\documentclass[manuscript,screen]{acmart}

\DeclareUnicodeCharacter{2060}{\nolinebreak}
\AtBeginDocument{%

  \providecommand\BibTeX{{%

\normalfont B\kern-0.5em{\scshape i\kern-0.25em b}\kern-0.8em\TeX}}}

\setcopyright{rightsretained}

\copyrightyear{2024}

\acmYear{2024}

\acmDOI{}

\acmConference[CCAI 2024]{CHI 2024 Workshop on LLMs as Research Tools: Applications and Evaluations in HCI Data Work}{May 12, 2024}{Honolulu, HI, USA}

\acmBooktitle{CHI 2024 Workshop on LLMs as Research Tools: Applications and Evaluations in HCI Data Work, May 12, 2024, Honolulu, HI, USA}

\AtBeginDocument{%
  \providecommand\BibTeX{{%
    \normalfont B\kern-0.5em{\scshape i\kern-0.25em b}\kern-0.8em\TeX}}}
\usepackage{booktabs}
\usepackage[normalem]{ulem}
\useunder{\uline}{\ul}{}
\begin{document}

\title{Leveraging Large Language Models (LLMs) to Support Collaborative Human-AI Online Risk Data Annotation}


\author{Jinkyung Park}
\affiliation{%
  \institution{Vanderbilt University}
  \city{Nashville}
  \country{USA}}
\email{jinkyung.park@vanderbilt.edu}


\author{Pamela Wisniewski}
\affiliation{%
  \institution{Vanderbilt University}
  \city{Nashville}
  \country{USA}}
\email{pamela.wisniewski@vanderbilt.edu}

\author{Vivek Singh}
\affiliation{%
  \institution{Rutgers University}
  \city{New Brunswick}
  \country{USA}}
\email{v.singh@rutgers.edu}


\begin{abstract}
In this position paper, we discuss the potential for leveraging LLMs as interactive research tools to facilitate collaboration between human coders and AI to effectively annotate online risk data at scale. Collaborative human-AI labeling is a promising approach to annotating large-scale and complex data for various tasks. Yet, tools and methods to support effective human-AI collaboration for data annotation are under-studied. This gap is pertinent because co-labeling tasks need to support a two-way interactive discussion that can add nuance and context, particularly in the context of online risk, which is highly subjective and contextualized. Therefore, we provide some of the early benefits and challenges of using LLMs-based tools for risk annotation and suggest future directions for the HCI research community to leverage LLMs as research tools to facilitate human-AI collaboration in contextualized online data annotation. Our research interests align very well with the purposes of the LLMs as Research Tools workshop to identify ongoing applications and challenges of using LLMs to work with data in HCI research. We anticipate learning valuable insights from organizers and participants into how LLMs can help reshape the HCI community’s methods for working with data.
\end{abstract}





\maketitle
\section{Introduction}

Online risk exposure is a pervasive phenomenon that affects millions of social media users every day ~\cite{han2018civility, hsueh2015leave, rosner2016verbal, zimmerman2016online}.  Given the massive scale of online content generation, the development and implementation of machine learning (ML) based risk detection tools to automatically identify and mitigate various online risks has been accelerating (e.g., ~\cite{sadeque2019incivility, stoll2020detecting, ozler2020fine, daxenberger2018automatically, kim2021you, park2023misinformation, park2023towards}). To develop these AI-based systems, Human-Computer Interaction (HCI), social computing, and ML research often employ a team of human coders to complete ground-truth data annotations (e.g., classifying whether messages are risky or not) through collaborative tasks among human annotators~\cite{muller2021designing, geiger2020garbage, barbosa2019rehumanized}, in which a coding scheme is pre-defined, a group of coders is trained with the scheme to reconcile disagreement, and inter-coder agreement is evaluated~\cite{o2020intercoder}.  
This often includes crowdsourced workers ~\cite{hosseinmardi2015analyzing}, a team of researchers and research assistants \cite{singh2017toward, park2023towards}, and/or domain experts ~\cite{park2023misinformation}. 
This annotation process involves an intensive and collaborative process of training, consensus-building, and quality control among multiple coders; therefore, it can be costly, time-consuming, and stress-inducing, while still yielding uneven levels of inter-coder agreement ~\cite{coe2014online, rains2017incivility}.  

Alternatively, proxies are used to label online risk collectively instead of annotating individual risk cases. For instance, existing literature on misinformation detection that relied on source-level labels assumes that all new articles from a given source share the same credibility level (reliable vs. unreliable) depending on the reputation of the source~\cite{horne2018assessing,grinberg2019fake, park2022toward}. However, in a recent study, researchers showed only about 50\% alignment between source level and article level labels for the credibility of political news articles and called for approaches to balance the quality and quantity of the ground truth annotation~\cite{park2023misinformation}. 
As such, there is a growing need for innovative and efficient methods to support human coders in labeling large corpora of online data, which can have a significant methodological impact on HCI and social computing research. In this position paper, we discuss the potential of the use of Large Language Models (LLMs)-based Conversational Agents (CAs) as AI-based co-coders for accurately annotating online risk data. Our position paper is highly relevant to the LLMs as Research Tools workshop as our focus on LLMs as research tools to facilitate online risk data annotation aligns very well with the core topic of interest for the workshop (i.e., LLMs to reshape the HCI community’s suite of methods for working with data).

\section{LLMs as Collaborative Agents for Online Risk Annotation}
After the recent release of various Large Language Model (LLM)-based Conversational Agents (CAs) (e.g., ChatGPT~\cite{OpenAI2023}), research communities are increasingly experimenting with data annotation tasks such as annotating political stance and sentiment of textual data~\cite{liu2023summary, zhang2022would, amin2023will, kuzman2023chatgpt}. Emerging literature suggests that LLM-based CAs can be useful for text classification tasks, owing to their accuracy and ability to flexibly adapt to tasks. 
For instance, Zhang et al.~\cite{zhang2022would} show that ChatGPT was able to annotate the political stance of the tweets with an average accuracy above 70. Similarly, Amin et al.~\cite{amin2023will} evaluated how accurately ChatGPT classifies sentiment, personality, and suicide ideation in given texts. The results showed that ChatGPT outperformed other robust language models for accurately classifying the sentiment of the text (i.e., positive, neutral, and negative classes). 
At the same time, challenges have been documented in the use of LLM-based CAs in annotating textual data for more contextualized constructs. For instance, the performance of ChatGPT for classifying the five personality and suicide ideation classifications was lower than the other pre-trained language models (e.g., BERT)~\cite{amin2023will}. The observations from prior studies indicate that ChatGPT is a generalist model that can perform many different text classification tasks without specialized training, yet dedicated training is required to achieve good results on specific tasks such as annotating contextualized online risk data.

Researchers have also explored the potential of LLM-based CAs as tools for identifying themes and grouping the textual data into identified themes (e.g., grounded thematic analysis) to reduce time and labor for such analysis~\cite{zhang2023qualigpt}. For instance, Zhang et al. developed GPT-based data analysis tools to identify major themes in textual data and highlighted the potential of leveraging LLMs for augmenting efficiency in qualitative data analysis. As such, prior studies highlighted the potential benefits of the use of LLMs as AI-based tools to facilitate text annotation tasks. 
However, most of the prior work shed light on the potential of LLM-based CAs for text annotation tasks to alternate human labor for the same tasks; hence, leaving potential for co-labeling tasks with human coders under-explored. This gap is pertinent because co-labeling tasks need to support a two-way interactive discussion that can add nuance and context, and help generate a rationale for the various decisions, particularly in the context of online risk behavior, which is highly contextual in nature~\cite{clay2003context}. 
While modern conversational agents have shown the ability to interact with humans and work with examples~\cite{mackeprang2019discovering, lai2022human}, their performance in collaborative text annotation exercises where different facets of co-labeling are important is as yet understudied. 
Here, we consider LLMs-based CAs as potentially useful co-labeling agents, which could support high-quality text annotations with explanations. Particularly, we situate this discussion in the data annotation of online risks. 

The concept of online risk is highly subjective and nuanced, and contextual sensitivity should be considered when identifying online risk~\cite{Litvinenko2023}. For instance, oftentimes, social media users use sarcasm or even more creative ways to get around content moderation~\cite{sadeque2019incivility}. This phenomenon has become increasingly common in user-generated content on social media platforms~\cite{rojas2017obstructing}. Hence, newer ways to innovate and re-think labeling in such settings could have a significant methodological impact on HCI and social computing research.  
To address the issues of annotating subjective and subtle online risk data using LLM-based CAs, Huang et al.~\cite{huang2023chatgpt} experimented with the ChatGPT to annotate hate speech in tweets into three categories: implicit hate speech, non-hate speech, and uncertain. The results showed that ChatGPT correctly recognized implicit hate speech in 80\% of the tweets with explanations better than those provided by human coders (i.e., crowdsourced annotators). The results from prior studies shed light on the potential of LLM-based CAs for text annotation tasks to alternate human labor for the same tasks. However, once again, it should be noted that the concept of risk and annotating online risk can be highly dependent on a variety of factors including cultural imprint, personality, political orientation, and contextual knowledge, hence, can lead to disagreement even among human coders. It is important that this insight does not lead to capitulation in the face of complexity but, instead, inspires better methods of automated analysis. Therefore, the combination of manual and automated content analysis is suggested as the gold standard for identifying online risk~\cite {esau2022content}. 
Here, we do not consider LLMs-based CAs as pre-trained classifiers for text annotation but rather as potentially useful co-labeling agents, which under the right settings of frequent and nuanced interaction could support high-quality annotations with explanations. Such an approach if feasible and successful, could allow for scalability in content analysis. Further, this can open doors to understanding nuances that might be lost even by human coders, who also have limited world awareness and cognitive abilities. 

\section{Design Considerations}
We consider LLM-based CAs as promising co-annotation partners because they are systems enabled with the ability to interact with the users using natural human dialogue ~\cite{rheu2021systematic} have shown promising results in text annotation tasks due to their accuracy and adaptability (e.g.,~\cite{zhang2023qualigpt, huang2023chatgpt,liu2023summary, zhang2022would, amin2023will, kuzman2023chatgpt}). The overarching question that we are interested in addressing is \textit{\textbf ``How can we design and implement a system that facilitates collaborative data annotation between researchers and conversational agents?''} 
We reflect on some of the potential challenges with the use of LLMs in contextualized and/or sensitive online risk data annotation tasks and provide design considerations for building LLMs-based research tools to effectively facilitate human-AI collaborative online risk data annotation. 

\begin{itemize}
    
\item \textbf{Interactivity}: Interactivity is the heart of LLMs. Partially, interaction between humans and AI through LLMs helps alleviate one of the major issues inherent to AI-based systems, transparency. To this end, “How can we leverage interactive conversation between humans and AI to support highly nuanced contextualized online risk data annotation tasks at scale?” 

\item \textbf{Context-Awareness}: LLMs work by tokenizing the texts into words and processing them~\cite{OpenAI_token_2023}. Therefore, how they understand risky interaction can differ from how humans do. The difference can be pertinent, especially for online risk data as risk is highly subjective and nuanced. Therefore, understanding how LLMs process textual data and make sense of information processed from textual data is a critical area for further exploration.

\item \textbf{Prompt Design}: The responses generated by LLMs-based models are highly dependent on the prompts. Therefore, the quality of prompts could be the foundation for effective human-AI collaboration for data annotation tasks. Recently, various prompting tuning approaches have been proposed and examined to elicit responses that are consistent and context-aware. Therefore, we ask “How can we design prompts that can best support highly nuanced and contextualized online risk data annotation tasks?”

\item \textbf{Consistency}: LLMs are known to provide inconsistent responses even with the same prompts given for the same query, and the reasons why they generate inconsistent responses are still in a black box. One way to increase consistency is through well-designed prompts given for specific tasks. Another option might be to combine responses from multiple iterations to identify consistent trends in the output. The remaining question here is “How can we design LLMs workflows to generate online risk labels with high consistency?” 

\item \textbf{User Interface}: There are potential features to maximize the benefit of LLMs-based online risk data annotation tools by helping (e.g.,  features to feed interaction logs between researchers and AI into prompts). At the same time, a complex user interface with advanced features may increase the burden for researchers to train themselves to use the systems. Moreover, it can make the data annotation process lengthy and prone to errors. Therefore, a streamlined user interface while providing a set of features to support effective human-AI collaboration for online risk data annotation tasks is needed.

\item \textbf{Data Privacy and Security}: When using LLMs to process online risk data, a primary concern for researchers is data privacy and security. 
For instance, OpenAI’s policies explicitly stipulate that data submitted by users through their API will not be used to train their models~\cite{OpenAI_security_2024}. Yet, we acknowledge that the conversation logs between researchers and AI can potentially be stored in the OpenAI API server for further usage. 
Therefore, researchers should also consider building LLM-based data annotation tools with private servers so that the training dataset is not shared via the web.

\end{itemize}

\section{conclusion}
Our research interests align very well with the purposes of the LLMs as Research Tools workshop to identify ongoing applications and challenges of using LLMs to work with data in HCI research. We anticipate learning more about organizers' and participants' ground-breaking research ideas to reshape the HCI community’s methods for working with data. In addition, participating in the workshop would be extremely beneficial for us to have a discourse on critical and ethical perspectives of the application of LLMs in HCI research. While we have identified some of the design condensations of LLM-based data annotation tools to support human-AI collaboration, we hope that participating in the worship will help us address some of the remaining challenges and come up with additional design implications. Finally attending the workshop will be a valuable opportunity to interact with and gain insights from organizers and participants, which could potentially lead to future collaboration opportunities.  

\section{About the Authors}

\textbf{Jinkyung Park} is a postdoctoral scholar in the Department of Computer Science at Vanderbilt University. Her research focuses on Human-Computer Interaction to promote online safety for youth and vulnerable populations.
\newline
\textbf{Pamela Wisniewski} is an associate professor in the Department of Computer Science at Vanderbilt University. Her work lies at the intersection of Human-Computer Interaction, Social Computing, and Privacy. Her expertise helps her empower end users and teach students to understand the value of user-centered design and evaluation.
\newline
\textbf{Vivek Singh} is an associate professor in the School of Communication and Information at Rutgers University. He designs AI systems that are responsive to human values and needs.



\bibliographystyle{ACM-Reference-Format}
\bibliography{00-References}


\begin{thebibliography}{38}


\ifx \showCODEN    \undefined \def \showCODEN     #1{\unskip}     \fi
\ifx \showDOI      \undefined \def \showDOI       #1{#1}\fi
\ifx \showISBNx    \undefined \def \showISBNx     #1{\unskip}     \fi
\ifx \showISBNxiii \undefined \def \showISBNxiii  #1{\unskip}     \fi
\ifx \showISSN     \undefined \def \showISSN      #1{\unskip}     \fi
\ifx \showLCCN     \undefined \def \showLCCN      #1{\unskip}     \fi
\ifx \shownote     \undefined \def \shownote      #1{#1}          \fi
\ifx \showarticletitle \undefined \def \showarticletitle #1{#1}   \fi
\ifx \showURL      \undefined \def \showURL       {\relax}        \fi
\providecommand\bibfield[2]{#2}
\providecommand\bibinfo[2]{#2}
\providecommand\natexlab[1]{#1}
\providecommand\showeprint[2][]{arXiv:#2}

\bibitem[Amin et~al\mbox{.}(2023)]%
        {amin2023will}
\bibfield{author}{\bibinfo{person}{Mostafa~M Amin}, \bibinfo{person}{Erik Cambria}, {and} \bibinfo{person}{Bj{\"o}rn~W Schuller}.} \bibinfo{year}{2023}\natexlab{}.
\newblock \showarticletitle{Will Affective Computing Emerge From Foundation Models and General Artificial Intelligence? A First Evaluation of ChatGPT}.
\newblock \bibinfo{journal}{\emph{IEEE Intelligent Systems}} \bibinfo{volume}{38}, \bibinfo{number}{2} (\bibinfo{year}{2023}), \bibinfo{pages}{15--23}.
\newblock


\bibitem[Barbosa and Chen(2019)]%
        {barbosa2019rehumanized}
\bibfield{author}{\bibinfo{person}{Nat{\~a}~M Barbosa} {and} \bibinfo{person}{Monchu Chen}.} \bibinfo{year}{2019}\natexlab{}.
\newblock \showarticletitle{Rehumanized crowdsourcing: A labeling framework addressing bias and ethics in machine learning}. In \bibinfo{booktitle}{\emph{Proceedings of the 2019 CHI Conference on Human Factors in Computing Systems}}. \bibinfo{pages}{1--12}.
\newblock


\bibitem[Clay-Warner(2003)]%
        {clay2003context}
\bibfield{author}{\bibinfo{person}{Jody Clay-Warner}.} \bibinfo{year}{2003}\natexlab{}.
\newblock \showarticletitle{The context of sexual violence: Situational predictors of self-protective actions}.
\newblock \bibinfo{journal}{\emph{Violence and victims}} \bibinfo{volume}{18}, \bibinfo{number}{5} (\bibinfo{year}{2003}), \bibinfo{pages}{543--556}.
\newblock


\bibitem[Coe et~al\mbox{.}(2014)]%
        {coe2014online}
\bibfield{author}{\bibinfo{person}{Kevin Coe}, \bibinfo{person}{Kate Kenski}, {and} \bibinfo{person}{Stephen~A Rains}.} \bibinfo{year}{2014}\natexlab{}.
\newblock \showarticletitle{Online and uncivil? Patterns and determinants of incivility in newspaper website comments}.
\newblock \bibinfo{journal}{\emph{Journal of Communication}} \bibinfo{volume}{64}, \bibinfo{number}{4} (\bibinfo{year}{2014}), \bibinfo{pages}{658--679}.
\newblock


\bibitem[Daxenberger et~al\mbox{.}(2018)]%
        {daxenberger2018automatically}
\bibfield{author}{\bibinfo{person}{Johannes Daxenberger}, \bibinfo{person}{Marc Ziegele}, \bibinfo{person}{Iryna Gurevych}, {and} \bibinfo{person}{Oliver Quiring}.} \bibinfo{year}{2018}\natexlab{}.
\newblock \showarticletitle{Automatically detecting incivility in online discussions of news media}. In \bibinfo{booktitle}{\emph{2018 IEEE 14th International Conference on e-Science (e-Science)}}. IEEE, \bibinfo{pages}{318--319}.
\newblock


\bibitem[Esau(2022)]%
        {esau2022content}
\bibfield{author}{\bibinfo{person}{Katharina Esau}.} \bibinfo{year}{2022}\natexlab{}.
\newblock \showarticletitle{Content Analysis in the Research Field of Incivility and Hate Speech in Online Communication}.
\newblock In \bibinfo{booktitle}{\emph{Standardisierte Inhaltsanalyse in der Kommunikationswissenschaft--Standardized Content Analysis in Communication Research: Ein Handbuch-A Handbook}}. \bibinfo{publisher}{Springer Fachmedien Wiesbaden Wiesbaden}, \bibinfo{pages}{451--461}.
\newblock


\bibitem[Geiger et~al\mbox{.}(2020)]%
        {geiger2020garbage}
\bibfield{author}{\bibinfo{person}{R~Stuart Geiger}, \bibinfo{person}{Kevin Yu}, \bibinfo{person}{Yanlai Yang}, \bibinfo{person}{Mindy Dai}, \bibinfo{person}{Jie Qiu}, \bibinfo{person}{Rebekah Tang}, {and} \bibinfo{person}{Jenny Huang}.} \bibinfo{year}{2020}\natexlab{}.
\newblock \showarticletitle{Garbage in, garbage out? Do machine learning application papers in social computing report where human-labeled training data comes from?}. In \bibinfo{booktitle}{\emph{Proceedings of the 2020 Conference on Fairness, Accountability, and Transparency}}. \bibinfo{pages}{325--336}.
\newblock


\bibitem[Grinberg et~al\mbox{.}(2019)]%
        {grinberg2019fake}
\bibfield{author}{\bibinfo{person}{Nir Grinberg}, \bibinfo{person}{Kenneth Joseph}, \bibinfo{person}{Lisa Friedland}, \bibinfo{person}{Briony Swire-Thompson}, {and} \bibinfo{person}{David Lazer}.} \bibinfo{year}{2019}\natexlab{}.
\newblock \showarticletitle{Fake news on Twitter during the 2016 US presidential election}.
\newblock \bibinfo{journal}{\emph{Science}} \bibinfo{volume}{363}, \bibinfo{number}{6425} (\bibinfo{year}{2019}), \bibinfo{pages}{374--378}.
\newblock


\bibitem[Han et~al\mbox{.}(2018)]%
        {han2018civility}
\bibfield{author}{\bibinfo{person}{Soo-Hye Han}, \bibinfo{person}{LeAnn~M Brazeal}, {and} \bibinfo{person}{Natalie Pennington}.} \bibinfo{year}{2018}\natexlab{}.
\newblock \showarticletitle{Is civility contagious? Examining the impact of modeling in online political discussions}.
\newblock \bibinfo{journal}{\emph{Social Media+ Society}} \bibinfo{volume}{4}, \bibinfo{number}{3} (\bibinfo{year}{2018}), \bibinfo{pages}{2056305118793404}.
\newblock


\bibitem[Horne et~al\mbox{.}(2018)]%
        {horne2018assessing}
\bibfield{author}{\bibinfo{person}{Benjamin~D Horne}, \bibinfo{person}{William Dron}, \bibinfo{person}{Sara Khedr}, {and} \bibinfo{person}{Sibel Adali}.} \bibinfo{year}{2018}\natexlab{}.
\newblock \showarticletitle{Assessing the news landscape: A multi-module toolkit for evaluating the credibility of news}. In \bibinfo{booktitle}{\emph{Companion Proceedings of the The Web Conference 2018}}. \bibinfo{pages}{235--238}.
\newblock


\bibitem[Hosseinmardi et~al\mbox{.}(2015)]%
        {hosseinmardi2015analyzing}
\bibfield{author}{\bibinfo{person}{Homa Hosseinmardi}, \bibinfo{person}{Sabrina~Arredondo Mattson}, \bibinfo{person}{Rahat~Ibn Rafiq}, \bibinfo{person}{Richard Han}, \bibinfo{person}{Qin Lv}, {and} \bibinfo{person}{Shivakant Mishra}.} \bibinfo{year}{2015}\natexlab{}.
\newblock \showarticletitle{Analyzing labeled cyberbullying incidents on the instagram social network}. In \bibinfo{booktitle}{\emph{International conference on social informatics}}. Springer, \bibinfo{pages}{49--66}.
\newblock


\bibitem[Hsueh et~al\mbox{.}(2015)]%
        {hsueh2015leave}
\bibfield{author}{\bibinfo{person}{Mark Hsueh}, \bibinfo{person}{Kumar Yogeeswaran}, {and} \bibinfo{person}{Sanna Malinen}.} \bibinfo{year}{2015}\natexlab{}.
\newblock \showarticletitle{“Leave your comment below”: Can biased online comments influence our own prejudicial attitudes and behaviors?}
\newblock \bibinfo{journal}{\emph{Human communication research}} \bibinfo{volume}{41}, \bibinfo{number}{4} (\bibinfo{year}{2015}), \bibinfo{pages}{557--576}.
\newblock


\bibitem[Huang et~al\mbox{.}(2023)]%
        {huang2023chatgpt}
\bibfield{author}{\bibinfo{person}{Fan Huang}, \bibinfo{person}{Haewoon Kwak}, {and} \bibinfo{person}{Jisun An}.} \bibinfo{year}{2023}\natexlab{}.
\newblock \showarticletitle{Is chatgpt better than human annotators? potential and limitations of chatgpt in explaining implicit hate speech}.
\newblock \bibinfo{journal}{\emph{arXiv preprint arXiv:2302.07736}} (\bibinfo{year}{2023}).
\newblock


\bibitem[Kim et~al\mbox{.}(2021)]%
        {kim2021you}
\bibfield{author}{\bibinfo{person}{Seunghyun Kim}, \bibinfo{person}{Afsaneh Razi}, \bibinfo{person}{Gianluca Stringhini}, \bibinfo{person}{Pamela~J Wisniewski}, {and} \bibinfo{person}{Munmun De~Choudhury}.} \bibinfo{year}{2021}\natexlab{}.
\newblock \showarticletitle{You Don't Know How I Feel: Insider-Outsider Perspective Gaps in Cyberbullying Risk Detection}. In \bibinfo{booktitle}{\emph{Proceedings of the International AAAI Conference on Web and Social Media}}, Vol.~\bibinfo{volume}{15}. \bibinfo{pages}{290--302}.
\newblock


\bibitem[Kuzman et~al\mbox{.}(2023)]%
        {kuzman2023chatgpt}
\bibfield{author}{\bibinfo{person}{Taja Kuzman}, \bibinfo{person}{Igor Mozetic}, {and} \bibinfo{person}{Nikola Ljube{\v{s}}ic}.} \bibinfo{year}{2023}\natexlab{}.
\newblock \showarticletitle{Chatgpt: Beginning of an end of manual linguistic data annotation? use case of automatic genre identification}.
\newblock \bibinfo{journal}{\emph{ArXiv, abs/2303.03953}} (\bibinfo{year}{2023}).
\newblock


\bibitem[Lai et~al\mbox{.}(2022)]%
        {lai2022human}
\bibfield{author}{\bibinfo{person}{Vivian Lai}, \bibinfo{person}{Samuel Carton}, \bibinfo{person}{Rajat Bhatnagar}, \bibinfo{person}{Q~Vera Liao}, \bibinfo{person}{Yunfeng Zhang}, {and} \bibinfo{person}{Chenhao Tan}.} \bibinfo{year}{2022}\natexlab{}.
\newblock \showarticletitle{Human-ai collaboration via conditional delegation: A case study of content moderation}. In \bibinfo{booktitle}{\emph{Proceedings of the 2022 CHI Conference on Human Factors in Computing Systems}}. \bibinfo{pages}{1--18}.
\newblock


\bibitem[Litvinenko(2023)]%
        {Litvinenko2023}
\bibfield{author}{\bibinfo{person}{Anna Litvinenko}.} \bibinfo{year}{2023}\natexlab{}.
\newblock \showarticletitle{The role of context in incivility research}.
\newblock In \bibinfo{booktitle}{\emph{Challenges and perspectives of hate speech research}}, \bibfield{editor}{\bibinfo{person}{Christian Strippel}, \bibinfo{person}{Sünje Paasch-Colberg}, \bibinfo{person}{Martin Emmer}, {and} \bibinfo{person}{Joachim Trebbe}} (Eds.). \bibinfo{series}{Digital Communication Research}, Vol.~\bibinfo{volume}{12}. \bibinfo{address}{Berlin}, \bibinfo{pages}{73--85}.
\newblock
\showISBNx{978-3-945681-12-1}
\showISSN{2198-7610}
\urldef\tempurl%
\url{https://doi.org/10.48541/dcr.v12.5}
\showDOI{\tempurl}


\bibitem[Liu et~al\mbox{.}(2023)]%
        {liu2023summary}
\bibfield{author}{\bibinfo{person}{Yiheng Liu}, \bibinfo{person}{Tianle Han}, \bibinfo{person}{Siyuan Ma}, \bibinfo{person}{Jiayue Zhang}, \bibinfo{person}{Yuanyuan Yang}, \bibinfo{person}{Jiaming Tian}, \bibinfo{person}{Hao He}, \bibinfo{person}{Antong Li}, \bibinfo{person}{Mengshen He}, \bibinfo{person}{Zhengliang Liu}, {et~al\mbox{.}}} \bibinfo{year}{2023}\natexlab{}.
\newblock \showarticletitle{Summary of ChatGPT-Related Research and Perspective Towards the Future of Large Language Models}.
\newblock \bibinfo{journal}{\emph{Meta-Radiology}} (\bibinfo{year}{2023}), \bibinfo{pages}{100017}.
\newblock


\bibitem[Mackeprang et~al\mbox{.}(2019)]%
        {mackeprang2019discovering}
\bibfield{author}{\bibinfo{person}{Maximilian Mackeprang}, \bibinfo{person}{Claudia M{\"u}ller-Birn}, {and} \bibinfo{person}{Maximilian~Timo Stauss}.} \bibinfo{year}{2019}\natexlab{}.
\newblock \showarticletitle{Discovering the sweet spot of human-computer configurations: A case study in information extraction}.
\newblock \bibinfo{journal}{\emph{Proceedings of the ACM on Human-Computer Interaction}} \bibinfo{volume}{3}, \bibinfo{number}{CSCW} (\bibinfo{year}{2019}), \bibinfo{pages}{1--30}.
\newblock


\bibitem[Muller et~al\mbox{.}(2021)]%
        {muller2021designing}
\bibfield{author}{\bibinfo{person}{Michael Muller}, \bibinfo{person}{Christine~T Wolf}, \bibinfo{person}{Josh Andres}, \bibinfo{person}{Michael Desmond}, \bibinfo{person}{Narendra~Nath Joshi}, \bibinfo{person}{Zahra Ashktorab}, \bibinfo{person}{Aabhas Sharma}, \bibinfo{person}{Kristina Brimijoin}, \bibinfo{person}{Qian Pan}, \bibinfo{person}{Evelyn Duesterwald}, {et~al\mbox{.}}} \bibinfo{year}{2021}\natexlab{}.
\newblock \showarticletitle{Designing ground truth and the social life of labels}. In \bibinfo{booktitle}{\emph{Proceedings of the 2021 CHI Conference on Human Factors in Computing Systems}}. \bibinfo{pages}{1--16}.
\newblock


\bibitem[OpenAI(2023)]%
        {OpenAI2023}
\bibfield{author}{\bibinfo{person}{OpenAI}.} \bibinfo{year}{2023}\natexlab{}.
\newblock \bibinfo{title}{Introducing ChatGPT}.
\newblock
\newblock
\urldef\tempurl%
\url{https://openai.com/blog/chatgpt}
\showURL{%
\tempurl}


\bibitem[OpenAI(2024a)]%
        {OpenAI_security_2024}
\bibfield{author}{\bibinfo{person}{OpenAI}.} \bibinfo{year}{2024}\natexlab{a}.
\newblock \bibinfo{title}{Security \& Privacy}.
\newblock
\newblock
\urldef\tempurl%
\url{https://openai.com/security}
\showURL{%
\tempurl}


\bibitem[OpenAI(2024b)]%
        {OpenAI_token_2023}
\bibfield{author}{\bibinfo{person}{OpenAI}.} \bibinfo{year}{2024}\natexlab{b}.
\newblock \bibinfo{title}{What are tokens and how to count them?}
\newblock
\newblock
\urldef\tempurl%
\url{https://help.openai.com/en/articles/4936856-what-are-tokens-and-how-to-count-them}
\showURL{%
\tempurl}


\bibitem[Ozler et~al\mbox{.}(2020)]%
        {ozler2020fine}
\bibfield{author}{\bibinfo{person}{Kadir~Bulut Ozler}, \bibinfo{person}{Kate Kenski}, \bibinfo{person}{Steve Rains}, \bibinfo{person}{Yotam Shmargad}, \bibinfo{person}{Kevin Coe}, {and} \bibinfo{person}{Steven Bethard}.} \bibinfo{year}{2020}\natexlab{}.
\newblock \showarticletitle{Fine-tuning for multi-domain and multi-label uncivil language detection}. In \bibinfo{booktitle}{\emph{Proceedings of the Fourth Workshop on Online Abuse and Harms}}. \bibinfo{pages}{28--33}.
\newblock


\bibitem[O’Connor and Joffe(2020)]%
        {o2020intercoder}
\bibfield{author}{\bibinfo{person}{Cliodhna O’Connor} {and} \bibinfo{person}{Helene Joffe}.} \bibinfo{year}{2020}\natexlab{}.
\newblock \showarticletitle{Intercoder reliability in qualitative research: debates and practical guidelines}.
\newblock \bibinfo{journal}{\emph{International journal of qualitative methods}}  \bibinfo{volume}{19} (\bibinfo{year}{2020}), \bibinfo{pages}{1609406919899220}.
\newblock


\bibitem[Park et~al\mbox{.}(2022)]%
        {park2022toward}
\bibfield{author}{\bibinfo{person}{Jinkyung Park}, \bibinfo{person}{Rahul Ellezhuthil}, \bibinfo{person}{Ramanathan Arunachalam}, \bibinfo{person}{Lauren Feldman}, {and} \bibinfo{person}{Vivek Singh}.} \bibinfo{year}{2022}\natexlab{}.
\newblock \showarticletitle{Toward Fairness in Misinformation Detection Algorithms}. In \bibinfo{booktitle}{\emph{Workshop Proceedings of the 16th International AAAI Conference on Web and Social Media. Retrieved from https://doi. org/10.36190}}.
\newblock


\bibitem[Park et~al\mbox{.}(2023a)]%
        {park2023misinformation}
\bibfield{author}{\bibinfo{person}{Jinkyung Park}, \bibinfo{person}{Rahul~Dev Ellezhuthil}, \bibinfo{person}{Joseph Isaac}, \bibinfo{person}{Christoph Mergerson}, \bibinfo{person}{Lauren Feldman}, {and} \bibinfo{person}{Vivek Singh}.} \bibinfo{year}{2023}\natexlab{a}.
\newblock \showarticletitle{Misinformation Detection Algorithms and Fairness across Political Ideologies: The Impact of Article Level Labeling}. In \bibinfo{booktitle}{\emph{Proceedings of the 15th ACM Web Science Conference 2023}}. \bibinfo{pages}{107--116}.
\newblock


\bibitem[Park et~al\mbox{.}(2023b)]%
        {park2023towards}
\bibfield{author}{\bibinfo{person}{Jinkyung Park}, \bibinfo{person}{Joshua Gracie}, \bibinfo{person}{Ashwaq Alsoubai}, \bibinfo{person}{Gianluca Stringhini}, \bibinfo{person}{Vivek Singh}, {and} \bibinfo{person}{Pamela Wisniewski}.} \bibinfo{year}{2023}\natexlab{b}.
\newblock \showarticletitle{Towards Automated Detection of Risky Images Shared by Youth on Social Media}. In \bibinfo{booktitle}{\emph{Companion Proceedings of the ACM Web Conference 2023}}. \bibinfo{pages}{1348--1357}.
\newblock


\bibitem[Rains et~al\mbox{.}(2017)]%
        {rains2017incivility}
\bibfield{author}{\bibinfo{person}{Stephen~A Rains}, \bibinfo{person}{Kate Kenski}, \bibinfo{person}{Kevin Coe}, {and} \bibinfo{person}{Jake Harwood}.} \bibinfo{year}{2017}\natexlab{}.
\newblock \showarticletitle{Incivility and political identity on the Internet: Intergroup factors as predictors of incivility in discussions of news online}.
\newblock \bibinfo{journal}{\emph{Journal of Computer-Mediated Communication}} \bibinfo{volume}{22}, \bibinfo{number}{4} (\bibinfo{year}{2017}), \bibinfo{pages}{163--178}.
\newblock


\bibitem[Rheu et~al\mbox{.}(2021)]%
        {rheu2021systematic}
\bibfield{author}{\bibinfo{person}{Minjin Rheu}, \bibinfo{person}{Ji~Youn Shin}, \bibinfo{person}{Wei Peng}, {and} \bibinfo{person}{Jina Huh-Yoo}.} \bibinfo{year}{2021}\natexlab{}.
\newblock \showarticletitle{Systematic review: Trust-building factors and implications for conversational agent design}.
\newblock \bibinfo{journal}{\emph{International Journal of Human--Computer Interaction}} \bibinfo{volume}{37}, \bibinfo{number}{1} (\bibinfo{year}{2021}), \bibinfo{pages}{81--96}.
\newblock


\bibitem[Rojas-Galeano(2017)]%
        {rojas2017obstructing}
\bibfield{author}{\bibinfo{person}{Sergio Rojas-Galeano}.} \bibinfo{year}{2017}\natexlab{}.
\newblock \showarticletitle{On obstructing obscenity obfuscation}.
\newblock \bibinfo{journal}{\emph{ACM Transactions on the Web (TWEB)}} \bibinfo{volume}{11}, \bibinfo{number}{2} (\bibinfo{year}{2017}), \bibinfo{pages}{1--24}.
\newblock


\bibitem[R{\"o}sner and Kr{\"a}mer(2016)]%
        {rosner2016verbal}
\bibfield{author}{\bibinfo{person}{Leonie R{\"o}sner} {and} \bibinfo{person}{Nicole~C Kr{\"a}mer}.} \bibinfo{year}{2016}\natexlab{}.
\newblock \showarticletitle{Verbal venting in the social web: Effects of anonymity and group norms on aggressive language use in online comments}.
\newblock \bibinfo{journal}{\emph{Social Media+ Society}} \bibinfo{volume}{2}, \bibinfo{number}{3} (\bibinfo{year}{2016}), \bibinfo{pages}{2056305116664220}.
\newblock


\bibitem[Sadeque et~al\mbox{.}(2019)]%
        {sadeque2019incivility}
\bibfield{author}{\bibinfo{person}{Farig Sadeque}, \bibinfo{person}{Stephen Rains}, \bibinfo{person}{Yotam Shmargad}, \bibinfo{person}{Kate Kenski}, \bibinfo{person}{Kevin Coe}, {and} \bibinfo{person}{Steven Bethard}.} \bibinfo{year}{2019}\natexlab{}.
\newblock \showarticletitle{Incivility detection in online comments}. In \bibinfo{booktitle}{\emph{Proceedings of the eighth joint conference on lexical and computational semantics (* SEM 2019)}}. \bibinfo{pages}{283--291}.
\newblock


\bibitem[Singh et~al\mbox{.}(2017)]%
        {singh2017toward}
\bibfield{author}{\bibinfo{person}{Vivek~K Singh}, \bibinfo{person}{Souvick Ghosh}, {and} \bibinfo{person}{Christin Jose}.} \bibinfo{year}{2017}\natexlab{}.
\newblock \showarticletitle{Toward multimodal cyberbullying detection}. In \bibinfo{booktitle}{\emph{Proceedings of the 2017 CHI Conference Extended Abstracts on Human Factors in Computing Systems}}. \bibinfo{pages}{2090--2099}.
\newblock


\bibitem[Stoll et~al\mbox{.}(2020)]%
        {stoll2020detecting}
\bibfield{author}{\bibinfo{person}{Anke Stoll}, \bibinfo{person}{Marc Ziegele}, {and} \bibinfo{person}{Oliver Quiring}.} \bibinfo{year}{2020}\natexlab{}.
\newblock \showarticletitle{Detecting impoliteness and incivility in online discussions: Classification approaches for German user comments}.
\newblock \bibinfo{journal}{\emph{Computational Communication Research}} \bibinfo{volume}{2}, \bibinfo{number}{1} (\bibinfo{year}{2020}), \bibinfo{pages}{109--134}.
\newblock


\bibitem[Zhang et~al\mbox{.}(2022)]%
        {zhang2022would}
\bibfield{author}{\bibinfo{person}{Bowen Zhang}, \bibinfo{person}{Daijun Ding}, {and} \bibinfo{person}{Liwen Jing}.} \bibinfo{year}{2022}\natexlab{}.
\newblock \showarticletitle{How would stance detection techniques evolve after the launch of chatgpt?}
\newblock \bibinfo{journal}{\emph{arXiv preprint arXiv:2212.14548}} (\bibinfo{year}{2022}).
\newblock


\bibitem[Zhang et~al\mbox{.}(2023)]%
        {zhang2023qualigpt}
\bibfield{author}{\bibinfo{person}{He Zhang}, \bibinfo{person}{Chuhao Wu}, \bibinfo{person}{Jingyi Xie}, \bibinfo{person}{ChanMin Kim}, {and} \bibinfo{person}{John~M Carroll}.} \bibinfo{year}{2023}\natexlab{}.
\newblock \showarticletitle{QualiGPT: GPT as an easy-to-use tool for qualitative coding}.
\newblock \bibinfo{journal}{\emph{arXiv preprint arXiv:2310.07061}} (\bibinfo{year}{2023}).
\newblock


\bibitem[Zimmerman and Ybarra(2016)]%
        {zimmerman2016online}
\bibfield{author}{\bibinfo{person}{Adam~G Zimmerman} {and} \bibinfo{person}{Gabriel~J Ybarra}.} \bibinfo{year}{2016}\natexlab{}.
\newblock \showarticletitle{Online aggression: The influences of anonymity and social modeling.}
\newblock \bibinfo{journal}{\emph{Psychology of Popular Media Culture}} \bibinfo{volume}{5}, \bibinfo{number}{2} (\bibinfo{year}{2016}), \bibinfo{pages}{181}.
\newblock


\end{thebibliography}




\end{document}